\documentclass{book}
\usepackage[contrib,lang=american]{ems-book} 

\usepackage{braket}

\newcommand{\beq}{\begin{equation}}
\newcommand{\eeq}{\end{equation}}

\DeclareMathOperator{\Tr}{Tr}
\DeclareMathOperator{\tr}{tr}
\DeclareMathOperator{\Ps}{P}
\DeclareMathOperator{\PS}{P_\text{sym}}

\begin{document}
\mainmatter

\title{Wehrl entropy, coherent states and quantum channels}
\titlemark{Wehrl entropy}

\emsauthor{1}{Peter Schupp}{P.~Schupp}


\emsaffil{1}{Jacobs University Bremen, Campus Ring 1, 28759 Bremen, Germany \email{p.schupp@jacobs-university.de}}

\classification[81R30]{81P45}


\begin{abstract}
We review Wehrl's definition of a semiclassical entropy in terms of coherent states and give an introductory overview of Lieb's conjecture, its proof (including earlier results), generalizations, and the role of covariant quantum channels in this context. These structures motivate an alternative definition of coherent states and have interesting physical applications and implications.\\[1em]
\emph{Dedicated to Elliott Lieb on the occasion of his 90th birthday.}
\end{abstract}

\makecontribtitle


\section{Wehrl entropy}

The quantum (von Neumann) entropy of a state described by a density matrix $\rho$,
\beq
S = - \tr \rho \ln \rho , \label{quantumS}
\eeq
is always non-negative, whereas the classical (Boltzmann) entropy for a continuous phase space distribution $\rho(q,p)$,
\beq
S_\text{cl} = - \int  \frac{dq dp}{2\pi\hbar} \, \rho(q,p) \ln \rho(q,p)  , \label{Gibbs}
\eeq
can in principle also take on arbitrarily negative 
values.\footnote{Throughout this article we shall use units, where Boltzmann's constant $k_B$ is equal to one and unless explicitly needed, e.g.\ in discussions of classical limits, we shall also set $\hbar =1$.} 
This seems to contradict the common expectation that classical expressions should arise from quantum mechanical ones in the limit of vanishing $\hbar$. Negative classical entropies can arise, because the integrand in~\eqref{Gibbs} is negative for values of  $\rho(q,p)$ that are larger than one. This can happen, if the classical distribution describes particles that are localized in phase space volumes that are smaller than $h = 2 \pi \hbar$, in contradiction with the quantum mechanical uncertainty relation. This situation should be rectifiable by a suitable smoothing of the classical phase space distribution. Arguing along these lines, Wehrl has suggested~\cite{Wehrl:1978zz} to replace the classical phase space distribution $\rho(q,p)$ in~\eqref{Gibbs}  by the expectation value of the quantum density matrix $\rho$ in coherent states of the harmonic oscillator, i.e.\ Gaussian minimum uncertainty states. The resulting semi-classical phase space distribution 
\beq
\rho(z) = \braket{ z | \rho | z},
\eeq 
is known as the as the Husimi distribution (or Q-function)~\cite{Husimi:1940} and also as the lower or covariant symbol of the density matrix $\rho$. It is the Weierstrass transform of the Wigner quasiprobability distribution, i.e. a smoothing by a Gaussian filter (as intended) and clearly $0 \leq \rho(z) \leq 1$. The entropy of $\rho(z)$ is called Wehrl entropy 
\beq
S_W = - \int dz \, \rho(z) \ln \rho(z) ,  \label{WehrlS}
\eeq
where we assume that the measure $dz$ is normalized such that $\int \rho(z)  \, dz = 1$. All distributions, density matrices and entropies that we have discussed so far can be defined on composite systems. Marginal distributions and respectively partial traces of density matrices are well defined and so is the entropy $S_{123}$ of the composite system, as well as the entropies $S_1$, $S_2$, $S_{12}$, $S_{23}$, etc.\ of the corresponding subsystems. It turns out that Wehrl entropy is very well behaved, see table~\ref{entropies}: It shares all the nice properties of the classical discrete Gibbs/Shannon  entropy $-\sum p_i \ln p_i$, even though the underlying distribution $\rho(z)$ of $S_W$ is continuous. We refer to~\cite{Liebconvexity:1975} for an expert overview of properties of quantum entropy. 

\begin{table}
\begin{tabular}{llccc}
\toprule
Entropy: & & quantum & Boltzmann  & Wehrl \\
\midrule
concavity & $S(\rho)$ concave in $\rho$ & \checkmark & \checkmark & \checkmark \\
positivity & $0 \leq S$ & \checkmark &  & \checkmark  \\
monotonicity & $S_1 \leq S_{12}$ &  &  & \checkmark \\
subadditivity & $S_{12} \leq S_1 + S_2$ & \checkmark &\checkmark & \checkmark \\
Araki-Lieb & $|S_1 -S_2| \leq S_{12}$ & \checkmark &  & \checkmark \\
strong subadditivity & $S_{123} + S_2 \leq S_{12} + S_{23}$ &  \checkmark & \checkmark & \checkmark \\
\bottomrule
\end{tabular} 
\caption{\label{entropies} Entropy properties and inequalities~\cite{Liebconvexity:1975,Wehrl:1978zz}.}
\end{table}

Using the properties of coherent states  (see next section), we can replace the trace in the definition of quantum entropy~\eqref{quantumS} by an integral over $z$, i.e. $S = - \int \bra z \rho \ln \rho \ket z \, dz$. Concavity of $ -x \ln x$  implies $S_W \geq S$. In fact, Wehrl showed $S_W > S$ and since $S \geq 0$, we find that even pure states have non-zero Wehrl entropy. The natural question arises, which states have minimal $S_W > 0$? By concavity (and uniqueness of the Fourier transform, see~\cite{Lieb:1978ys}), these must be pure states and Wehrl conjectured that $S_W \geq 1$, where the minimum is attained for coherent states. That is Wehrl's conjecture~\cite{Wehrl:1979} -- it was proven by Lieb in~\cite{Lieb:1978ys}.\footnote{A curious fact is that Lieb's proof actually appeared before Wehrl's conjecture.} More precisely, Lieb proved a more general inequality for a  
R\'enyi-type Wehrl entropy, where $-x\ln x$ in~\eqref{WehrlS} is replaced by $x^s/(1-s)$; the original entropy is recovered in the limit $s\rightarrow 1$.  Lieb's proof is based on the strengthened Hausdorff-Young inequality~\cite{Beckner:1975} and the sharp Young inequality~\cite{Beckner:1975,BRASCAMP1976151}. Both have Gaussian optimizers and so does Wehrl's inequality. Uniqueness of the minimizers was proven by Carlen~\cite{Carlen1991SomeII}. For an alternative proof based on hypercontractivity see~\cite{Luo_2000}. The minimal Wehrl entropy for fixed von Neumann entropy was studied in~\cite{DePalma2018}.

While Lieb's proof is rather slick, it is surprising that such deep results from harmonic analysis were needed. In order to get a better, perhaps more group-theoretic understanding, Lieb  suggested to study the analog of Wehrl's conjecture for spin coherent states, as a finite dimensional and thus hopefully simpler problem.  This is Lieb's conjecture.  It is well known in the mathematical physics community and many people tried to solve it, but it remained open for over thirty years. Some partial results were found in the meantime: Coherent states were shown to be a shallow local minimum in~\cite{Lee_1988}, they were shown to be unique minimizers for spin 1 and 3/2 as well as for all integer R\'enyi entropies for all spin  in~\cite{Schupp1999OnLC}, for spin 1 this was also shown independently in~\cite{Horia2002},  sharp high spin asymptotics were settled in~\cite{Bodman2004}. The conjecture was finally settled by Lieb and Solovej in~\cite{lieb2014proof} and further generalized in~\cite{lieb2016proof} and~\cite{lieb2019wehrl}. The uniqueness of the minimizers for spin greater than $3/2$ is however still open. In the following sections we will give an overview of the problem, the results and generalizations.  For more background information and further details on all topics in this article, we refer the reader to the excellent book~\cite{bengtsson_zyczkowski_2017} by Bengtsson and Życzkowski and of course to the original articles.

\section{Coherent states}

Looking for quantum states that are as classical as possible, Schrödinger introduced coherent states~$\ket z$ as displacements of the minimum uncertainty ground state~$\ket 0$ of the quantum harmonic oscillator in spatial as well as momentum direction~\cite{1926Schroedinger} -- in analogy to the  initial spatial displacement and momentum of a classical oscillator like a pendulum. For harmonic oscillator coherent states, the displacement is generated by the Heisenberg group and is labeled by $z = \frac 1{\sqrt 2}(q+ip)$. The ground state $\ket 0$ is not only an energy eigenstate, but also an eigenstate of the lowering operator $a$ with eigenvalue zero. The latter property is inherited by the displaced lowering operator and ground state, i.e.\ $(a-z) \ket z = 0$, which  is usually written as: $a \ket z = z \ket z$ and can be used as an alternative definition  for these Schrödinger/Klauder/Glauber coherent states. 

Coherent states $\ket z$ are thus elements of the orbit of the ground state $\ket 0$  under the Heisenberg group. This notion can be generalized to orbits of a fiducial vector in some representation of a Lie group under the action of that group~\cite{Perelomov:book}. The choice of the fiducial vector is essential for the properties of the resulting coherent states:  It should be a state of a enhanced symmetry (which may not be obvious without complexification.) 
For compact Lie groups, highest weight vectors are such states of enhanced symmetry: They are eigenstates of all generators in the Borel subalgebra of the corresponding complexified Lie algebra. (They are eigenstates with eigenvalue zero of the raising operators.) 
Keeping the notation $\ket z$ also for the generalized coherent states, we shall collect some key properties: Coherent states satisfy a completeness relation
\beq
\int dz \, | z\rangle\!\langle z| = \mathbf 1 , \label{complete-z}
\eeq
which implies $\tr A = \int dz \, \langle z |A| z \rangle$ (with suitably normalized measure $dz$), but they are not orthogonal, i.e.\ they form a so-called overcomplete basis. More precisely, we are dealing with a coherent-state positive operator-valued measure (POVM).

A striking property of coherent states is that the diagonal matrix elements (lower symbol)
\beq
A( z) = \langle z |A| z \rangle 
\eeq
of an operator $A$ (typically) already determine that operator uniquely: Let $C = A - B$ with a second operator $B$, then $C( z) = 0$ for all $ z$ implies $C = 0$, i.e. $A=B$. The proof uses analytic properties of the lower symbol.  The lower symbol is thus a faithful representation of an operator. 
Another interesting property is that any operator $A$  can be expanded diagonally in coherent states
\beq
A = \int d  z \, h_A( z) | z \rangle\!\langle z |  , \label{upper}
\eeq
where $h_A( z)$ is called an upper symbol of $A$ (upper symbols are not unique). These two properties are in fact closely related: Contracting (\ref{upper}) with an operator $C$ gives
$\Tr (C^\dagger A) \propto \int d  z \,  \overline{C(z)} \, h_A( z)$, i.e. the operators that can be represented by an upper symbol as in  (\ref{upper}), are orthogonal to the operators that are in the kernel of the lower symbol map.  Hermitean operators have real lower und upper symbols. Positive semidefinite operators and density matrices have unique non-negative lower symbols, but the same is in general not true for upper symbols. 

Following Wehrl, these properties suggest to interpret the lower symbol of a density matrix $\rho$, which is by definition positive semidefinite and normalized, as a probability density.
This is also natural from the coherent-state POVM measurement point of view: $\rho(z) = \tr( \rho  | z \rangle\!\langle z | )$ is precisely the probability density for a measurement of $z$. Consider now $ \int d  z \,  \phi( \rho( z))$ for various functions $\phi$: For $\phi(x) = x$ we can verify the normalization and get $\tr \rho = 1$. With $\phi(x) = -x \ln x$ we obtain the Wehrl entropy. As already pointed out by Lieb~\cite{Lieb:1978ys}, the entropy conjecture is trivially true  for density matrices that can be expressed in terms of a non-negative upper symbol. For coherent states this upper symbol is a delta function, but pure states can unfortunately in general not be written in terms of non-negative upper symbols. 
One can also choose other functions for~$\phi$, e.g. $\phi(x) = x^s/(1-s)$, which gives the R\'eyni-
Wehrl entropy, or $\phi(x) = x (1-x)$, which gives a quadratic approximation to entropy. Defining  entropy more generally relative to a POVM $\sum_n E_n = \mathbf 1$, $E_n \geq 0$, as $S = - \sum_n p_n \ln p_n$ with $p_n = \tr (\rho E_n)$, we see that the Wehrl entropy is obtained for a coherent-state POVM, while the usual quantum von Neumann entropy is obtained for an eigen-POVM of $\rho$ and is in fact the minimum of all such entropies.

The lower symbol of a product $A B$ of operators can be written in terms of a star product $\star$ of the lower symbols of $A$ and $B$, 
\beq
A(z) \star B(z) = \langle z | A B | z \rangle ,
\eeq
i.e.\ a formal power series in $\hbar$ of bidifferential operators acting on the functions $A(z)$ and $B(z)$, such that $\star$ is associative. That is a starting point for a phase space formulation of quantum mechanics. For the original Schrödinger coherent states, it yields the Wick-Voros star product and corresponds to a normal-ordered quantization prescription. Star-versions of functions can be defined in analogy to the definition of functions of operators. The quantum mechanical entropy of a density matrix can then at least formally be written in terms of the lower symbol as
\beq
S = - \int dz \, \rho(z) \star \ln_\star \rho(z) .
\eeq
At zeroth order in $\hbar$, i.e.\ ``classically'',  this expression gives the Wehrl entropy~\eqref{WehrlS}. (This is of course not meant as a formal proof of a classical limit of the quantum entropy, but rather as further motivation.) Later we shall see that the ``classical'' Wehrl entropy is in fact a quantum entropy, namely that of a density matrix observed through a certain covariant quantum channel in a suitable limit.

\subsection{Spin coherent states}

Spin coherent states -- also called Bloch coherent states -- in a spin-$l$ irreducible representation $[l] \equiv {\mathbb C}^{2l+1}$ of $SU(2)$ with $2l+1 \in \mathbb N$
are defined as orbits of the highest weight vector $|l,l\rangle$. The stability group of that vector  is $U(1)$ and  spin coherent states can thus be labeled by points $\Omega =(\theta, \phi)$ on the sphere $S_2 \cong SU(2)/U(1)$,
\beq
|\Omega_l\rangle = \mathcal R(\Omega) |l,l\rangle = \sum_{m=-l}^l \begin{pmatrix} 2l \\ l+m \end{pmatrix}^{\!\frac 12} e^{-im \phi /2}\,\cos^{l+m}(\tfrac\theta 2) \, \sin^{l-m}(\tfrac\theta 2) \, |l,m\rangle ,
\eeq
where $\mathcal R(\Omega)$ denotes a rotation that takes the north pole to the point $\Omega$ and $l$ labels the representation of $SU(2)$. The irreducible representations of $SU(2)$ are symmetric, i.e.\ they have single-row Young tableaux, and can be written as symmetrized tensor products of spin-$\frac 12$ representations. A generic spin-$l$ state can thus be written as the projection $\Ps_l$ onto the fully symmetric part, i.e. onto the spin-$l$ representation $[l]$ of the tensor product of $2l$ spin-$\frac 12$ states, 
\beq 
|\psi \rangle = c_\psi \Ps_l |\omega_1 \otimes \ldots \otimes \omega_{2l}\rangle ,  \label{stelarrep}
\eeq
where the $\omega_i = (\theta_i,\psi_i)$ denote unit vectors up to a phase in $\mathbb C^2$, i.e.\ points on the Bloch sphere $\mathbb {CP}^1 \cong S_2$ that parametrize spin-$\frac 12$ states and $c_\psi$ is a normalization constant. This is also known as the stellar representation (points on the Bloch sphere $\sim$ stars in the sky) and can be used for a fast computation of multipole vectors~\cite{Helling:2006xh}. For spin coherent states, no projection is needed: They are already fully symmetric  $|\Omega_l\rangle = |\Omega \otimes \ldots \otimes \Omega\rangle$ and the tensor product of coherent states is again a coherent state:
\beq
|\Omega_l\rangle \otimes |\Omega_{j}\rangle 
=  |\Omega_{l+j}\rangle. \label{productproperty}
\eeq
Spin coherent states are complete via Schur's lemma
\beq
(2l +1) \int \frac{d \Omega}{4\pi}\, |\Omega_l\rangle\!\langle\Omega_l| =  \Ps_l  , \label{complete}
\eeq
where $\Ps_l$ is the projector onto $[l]$. They are normalized $\langle\Omega_l|\Omega_l\rangle = 1$ but not orthogonal
\beq
|\langle \Omega_l | \Omega'_l\rangle|^2 = \cos^{4l} (\mbox{\large$\sphericalangle$}{(\Omega, \Omega')})  , \label{notorthogonal}
\eeq
i.e.\ they form an overcomplete basis of $[l]$ (a spin-coherent-state POVM actually).
In the $l \rightarrow \infty$ limit, $(2l+1) |\langle \Omega_l | \Omega'_l\rangle|^2$ becomes a delta function $\delta(\Omega, \Omega')$  and in this limit the coherent states form an infinite-dimensional orthonormal basis labeled by points on the sphere.

The Lieb- Wehrl entropy of spin coherent states is
\beq
S_W(\rho) = -(2l +1) \int \frac{d \Omega}{4\pi} \, \rho(\Omega) \ln \rho(\Omega).  \label{Wehrl}
\eeq
For coherent states $S_W = \frac{2j}{2j+1}$ and according to Lieb's conjecture, proven in~\cite{lieb2014proof}, this is the minimum value for all states. 

The stellar representation factorizes $\rho(\Omega)$, thus turning the logarithm term into a sum and allowing an explicit computation of the Wehrl entropy, leading to nice geometric expressions in terms of symmetric polynominals in the cordal distances between points on the Bloch sphere~\cite{Schupp1999OnLC}. For spin~1 the Wehrl entropy is given by 
\beq
S_W = \frac 23 + c \cdot \left(\frac\mu 2 + \frac 1c \ln \frac 1c \right)  \quad \text{with} \quad \frac 1c = 1 - \frac \mu 2,
\eeq
where $\mu$ is the square cordal distance between two points on the Bloch sphere that define the underlying state.
For spin 3/2 the Wehrl entropy is
\beq
S_W = \frac 34 + c \cdot \left( \frac{\epsilon + \mu + \nu}{3} - \frac{\epsilon \mu + \epsilon \nu + \mu \nu}6  + \frac 1c \ln \frac 1c \right)  \quad \text{with} \quad \frac 1c = 1- \frac{\epsilon + \mu + \nu}{3} ,
\eeq
where $\epsilon$, $\mu$, $\nu$ are the square cordal distances between three points on the Bloch sphere that define the state.  Coherent states are the unique states for which the cordal distances vanish, thus minimizing the Wehrl entropies~\cite{Schupp1999OnLC}. For other states, similar expressions can be found and Weingart has managed to determine them in closed form for higher values of spin~\cite{Schupp1999OnLC,bengtsson_zyczkowski_2017,Weingart}.

In order to compute (integer) R\'enyi entropies, one  needs to replace $-\rho(\Omega) \ln \rho(\Omega)$ in \eqref{Wehrl} by $\left(\rho(\Omega)\right)^n$. The resulting integral yields the magnitude of the projection onto the completely symmetric (maximum spin) part of $\rho^{\otimes n}$, i.e.\ $\tr \Ps_{n\cdot j}(\rho^{\otimes n})$ (up to a positive factor). In view of~\eqref{productproperty} it is easy to see that the unique maximizers of this expression are coherent states~\cite{Schupp1999OnLC}. By a similar argument one finds that this also holds for symmetric $SU(N)$ coherent states, i.e.\ they are the \emph{unique} extremizers of all integer R\'enyi entropies. 

Based on theoretical arguments and extensive numerical experiments, it pretty soon became clear that the Wehrl-Lieb conjecture should not only hold for Shannon or R\'enyi-type entropies, but quite generally for any concave (or convex) function~\cite{SchuppKolloquium2004}. The key idea to solving Lieb's conjecture for all spin turns out to be a further generalization, namely to replace the map $\rho \mapsto \rho(\Omega)$ by a suitable covariant quantum channel, i.e. by a trace-preserving completely positive map that commutes with the adjoint action of the underlying symmetry group (here: $SU(2)$ and later $SU(N)$). It can be proven that the image of coherent states under this quantum channel majorizes the image of all other states. In the infinite dimensional limit the eigenvalues of the resulting matrices approach the values of the lower symbol and the conjecture follows. 

In the following section we will first present a toy model (that may be of interest in its own right) and then show how to reformulate Lieb's conjecture in terms of quantum channels. Here we follow an approach (Schupp 2008 unpublished and~\cite{PhysRevD.99.103501}) that makes it fairly easy to see how the quantum coherent operators (covariant quantum channels) introduced in \cite{Lieb:1991np} arise that were then eventually used in the proof of the conjecture and its generalizations~\cite{lieb2014proof,lieb2016proof}. Given the quantum channel we then sketch the beautiful proof of Lieb and Solovej for symmetric $SU(N)$ coherent states.

\section{Covariant quantum channels}

Covariant quantum channels are completely positive trace-preserving maps $\Phi$ between linear operators on Hilbert spaces $\mathcal H_1$, $\mathcal H_2$ that are covariant with respect to a symmetry group $G$, i.e.\ $\Phi \big(U_1(g) {\rho U_1}^\dagger(g) \big)= U_2(g) \Phi (\rho) {U_2}^\dagger(g)$ for all $g \in G$, where $U_1$ and $U_2$ are unitary representations of $G$ on $\mathcal H_1$ and $\mathcal H_2$ respectively. Here we shall focus on $SU(2)$ and more generally $SU(N)$ and consider only unital maps. The quantum (von Neumann/Shannon) entropy of the image of a density matrix under any one of these maps defines a new ``covariant'' mixing entropy $S(\Phi(\rho))$   that shares many of the nice properties of the already mentioned entropies. In particular these covariant entropies are strictly larger than the original quantum entropy (even for pure states), provided that the quantum channel is not just a simple unitary transformation. The question arises, which states minimize the covariant entropies and natural candidates are coherent states of the underlying symmetry group. A particular type of these quantum channels is in fact directly related to the Wehrl entropy in a certain limit as we shall see. 
There is actually no need to consider only Shannon-type entropies -- one can consider general concave (or convex) functions and similar inequalities will hold. More generally one should study matrix majorization of the quantum channel images of states. Recall that a matrix majorizes another one with equal trace, if all partial sums of the largest eigenvalues of the first matrix are at least as large as the corresponding sums for the second matrix. Inequalities for Schur-concave (or convex) functions follow from this. 

In fact this approach could be turned around and leads to a proposal for a novel definition of coherent states based on quantum channels, namely states whose image under a given quantum channel (or a class of quantum channels) majorize the channel images of all other states. An equivalent more geometric formulation in terms of extreme points of convex polytopes along the lines of a generalized Schur-Horn theorem with two independent unitary orbits (one on the input, one on the output of the quantum channel) is also possible. Right now such a new definition of coherent states is still of limited practical use (except that it would conveniently turn difficult to proof entropy conjectures into tautologies.) But once we have more Lieb-Solovej-type theorems and a better understanding of the underlying mathematics, it could become a powerful tool: It would do away with ambiguities in the choice of fiducial vectors and it would generalize the concept of symmetry groups underlying the current definition of coherent states. From the new point of view, coherent states would be considered to be the ``purest'' among all pure states. The new definition will also be more physical: It answers the question, into which ``classical'' states a system will likely collapse, when observed in an in-perfect way modeled by a quantum channel, namely into one of the majorizing coherent states.  This has the potential to give a mathematical rigorous explanation for the fundamental question, why the world looks classical. 

Here is a ``toy model'' of a covariant quantum channel: Let $L_1$, $L_2$, $L_3$ be the standard angular momentum generators in the $2j+1$-dimensional spin-$l$ representation and define a quantum channel and ``angular'' entropy via
\beq
\rho \mapsto \rho_\text{ang} = \frac{1}{l(l+1)} \sum_{i=1}^3 L_i \rho L_i{}^\dagger  , \quad S_\text{ang} = - \Tr \rho_\text{ang} \ln \rho_\text{ang} . \label{angentropy}
\eeq
The transformation is obviously of Kraus form and therefore completely positive. It is trace-preserving, unital and covariant, i.e. it commutes with the unitary $SU(2)$ transformations, because $C = \sum_i L_i{}^\dagger L_i$ is the quadratic casimir and has value $l(l+1)$ in the spin $l$ representation. The formula for angular entropy can be written in a basis-independent way by replacing 
$\sum L_i \otimes L_i$ by $\tfrac 12 (\Delta C - C\otimes 1 - 1 \otimes C)$, where $\Delta C$  the coproduct of the casimir. In practice the formula is usually rewritten in terms of $\tfrac 1{\sqrt 2} L_\pm$ instead of $L_1$ and $L_2$. Therefore we have included the dagger $\dagger$ in (\ref{angentropy}), which is of course not necessary for hermitean $L_i$.
For a pure state $\rho = |\psi\rangle\!\langle \psi|$, there is also a dual Gram matrix formulation of the angular entropy:
\beq
G_{ij} = \langle \psi | C^{-1} L_i{}^\dagger L_j|\psi\rangle , \quad S_\text{ang} = - \Tr ( G \ln G ) . \label{anggram}
\eeq
Recall that the Gram matrix has the same non-zero singular (eigen) values as the original matrix.  For a coherent state, we find the eigenvalue tupel $(j^2,j,0)$. For low values of $j$, it is not too hard to show that the coherent state eigenvalue tupel majorizes the corresponding eigenvalue tuple for any other state and we hence get the desired entropy inequalities.
%

\subsection{Projection entropy}

\begin{figure}[t]
\includegraphics[width=.6\textwidth]{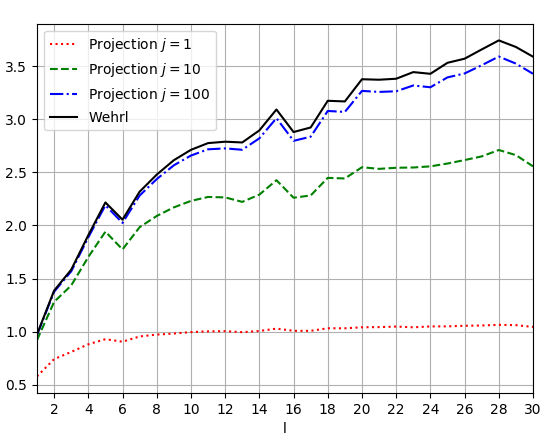}
\caption{Wehrl entropy versus $j = 1, 10, 100$ projection entropies for integer spin $l$ states $\ket {\psi_l} = \sum_m a_{lm}\ket{l,m}$, used as a tool in the analysis of cosmic microwave background data~\cite{PhysRevD.99.103501}.}\label{plots}
\end{figure}

Let us return to the Wehrl entropy of spin coherent states; closely following~\cite{PhysRevD.99.103501,schupp2008}, we shall see how it is related to the  quantum coherent operators (covariant quantum channels) introduced in \cite{Lieb:1991np}.
Let $\rho$ be a density matrix on  $[l] = \mathbb C^{2l+1}$  and introduce an ancilla Hilbertspace $[j] = \mathbb C^{2j+1}$. Using the product property (\ref{productproperty}) and normalization of coherent states, we can rewrite the lower symbol $\rho(\Omega)$ that enters the formula for the Wehrl entropy as follows:
\beq
\langle\Omega_l|\rho|\Omega_l\rangle = \langle\Omega_l|\rho|\Omega_l\rangle \langle\Omega_j|\Omega_j\rangle
= \langle\Omega_l \otimes \Omega_j|\rho \otimes \mathbf 1|\Omega_l \otimes \Omega_j\rangle
= \langle\Omega_{l+j}|\rho \otimes \mathbf 1|\Omega_{l+j}\rangle \ ,
\eeq
where  $\mathbf 1$ is the unit operator on $[j]$. The values of the lower symbol are thus the diagonal elements of a family of infinite-dimensional matrices
\beq
\rho_j(\Omega, \Omega') = \langle\Omega_{l+j}|\rho \otimes \mathbf 1|\Omega'_{l+j}\rangle \ . \label{infdimmatrix}
\eeq
By an infinite-dimensional compact analog of the Schur-Horn theorem the diagonal elements $\rho(\Omega)$ are majorized by the eigenvalues of the $\rho_j(\Omega, \Omega')$ matrices. This implies that any concave function of the values $\rho(\Omega)$ will be larger than or equal to the respective function of the eigenvalues of $\rho_j(\Omega, \Omega')$. The Wehrl entropy is therefore larger than or equal to the von Neumann entropy of $\rho_j(\Omega, \Omega')$. For convex functions the inequalities are reversed.  In the limit $j \rightarrow \infty$ and  in view of (\ref{notorthogonal}), the off-diagonal matrix elements of $\rho_j(\Omega, \Omega')$ become zero and the inequalities become equalities. 
Using (\ref{complete}) on both sides of (\ref{infdimmatrix}) we obtain a finite-dimensional matrix 
\beq
\Ps_{l+j}\left(\rho \otimes \mathbf 1\right) \Ps_{l+j} \label{projmatrix}
\eeq
from $\rho_j(\Omega, \Omega')$, where $\Ps_{l+j}$ is the projector onto the highest spin component $[l+j]$ of the tensor product. The matrix (\ref{projmatrix}) has the same eigenvalues as  $\rho_j(\Omega, \Omega')$ in the following sense:
\beq
\Ps_{l+j}\left(\rho \otimes \mathbf 1\right) \Ps_{l+j} |V_\lambda\rangle = \lambda |V_\lambda\rangle \label{eigenvalproj}
\eeq
implies that $V_\lambda (\Omega) := \langle \Omega_{l+j} | V_\lambda\rangle$ satisfies
\beq
(2(l+j) + 1) \int \frac{d \Omega'}{4\pi} \,  \langle\Omega_{l+j}|\rho \otimes \mathbf 1|\Omega'_{l+j}\rangle V_\lambda(\Omega') = \lambda V_\lambda(\Omega) \label{eigenvalint}
\eeq
and vice versa if $V_\lambda(\Omega)$ is a solution of (\ref{eigenvalint}) then 
\[
|V_\lambda\rangle = (2(l+j) +1) \int \frac{d \Omega}{4\pi} \, |\Omega_{l+j}\rangle V_\lambda (\Omega)
\] 
satisfies (\ref{eigenvalproj}). We have thus found that the eigenvalues of the matrix (\ref{projmatrix}) majorize the values $\rho(\Omega)$ of the lower symbol of $\rho$ in the sense explained above, namely that inequalities are implied for concave (or convex) functions of these values. It can be seen with a simple convexity argument that pure states majorize mixed ones and we shall see that among the pure states, projectors $|\Omega\rangle\!\langle \Omega|$ onto coherent states will lead to matrices (\ref{eigenvalproj}) that majorize all other choices.
Among the concave functionals we are in particular interested in  entropy and define an appropriately normalized mixed density matrix
\beq
\rho_{\text{pro},j} = \frac{2l+1}{2(l+j)+1} \Ps_{l+j}\left(\rho \otimes \mathbf 1\right) \Ps_{l+j} \ , \label{mixproj}
\eeq
whose von Neumann entropy is what we call the ``projection entropy'' 
\beq
S_{\text{pro},j}(\rho) = \Tr \phi\Big(\frac {2l+1}{2(l+j)+1} \Ps_{l+j}\big(\rho\otimes \mathbf 1\big)\Ps_{l+j}\Big) \qquad \phi(x) \equiv -x \ln x \ .
\eeq
From the fact that the mixed density matrix (\ref{mixproj}) has at most $2j+1$ non-zero eigenvalues, we get an upper bound for the projection entropy: $S_{\text{pro},j}(\rho)  \leq \ln(2j+1)$.
From the $[l+j]$-perspective the Wehrl entropy should also  be computed from (\ref{mixproj}) and we get the aforementioned inequalities. The only difference from the original definition of Wehrl entropy (\ref{Wehrl}) is a rescaling of the density matrix and related renormalization of the integral, which leads to a shift in entropy and the following inequality:
\beq
S_W(\rho) \geq S_{\text{pro},j}(\rho) + \ln\left( \frac {2l+1}{2(l+j)+1}\right)  .
\eeq
In the limit $j \rightarrow \infty$ this inequality becomes an equality, see figure~\ref{plots} for an illustration. 
The projector $\Ps_{l+j}: [l] \otimes [j] \rightarrow [l+j]$ can be expressed in terms of Clebsch-Gordan coefficients and more elegantly in a second quantized formulation that is then also used to prove the conjectures. 
For large $j$ the projection method provides a good way to compute the Wehrl entropy with high precision. For small $j$ we get an entropy measure with  the nice properties of Wehrl entropy, but a pretty large computational advantage. Let us consider the case where $\rho$ is a pure state, i.e. $\rho = |\psi_l\rangle\!\langle\psi_l|$. For a pure state the matrix (\ref{projmatrix}) can be rewritten as the Gram matrix of a set of vectors $\vec V_M \in [l+j]$ that are labeled by a basis of $[j]$:
\beq
\Ps_{l+j}\left(|\psi_l\rangle\!\langle\psi_l| \otimes \mathbf 1\right) \Ps_{l+j} = \sum_{M=-j}^j \vec V_M \vec V_M{}^\dagger \ , \qquad \vec V_M = \Ps_{l+j}\big(|\psi_l\rangle\otimes|j,M\rangle\big) \ .
\eeq
The dual Gram matrix
\beq
\Tr_{[l+j]} \big( \vec V_M \vec V_{M'}{}^\dagger \big) = \vec V_{M'}{}^\dagger  \cdot  \vec V_M =  
 \big(\langle\psi_l|\otimes\langle j,M'|\big) \Ps_{l+j} \big(|\psi_l\rangle\otimes|j,M\rangle\big)
\eeq
has the same non-zero eigenvalues as the original matrix, because for any matrix $C$, $C C^\dagger$ and $C^\dagger C$ have the same non-zero singular values. 
We can therefore also use the dual Gram matrix for the computation of the projection entropy. Appropriately normalized and written in basis-independent notation we have
\beq
\tilde \rho_{\text{pro},j} = \frac{2l+1}{2(l+j)+1} \langle \psi_l \otimes \text{id}| \Ps_{l+j}|\psi_l \otimes \text{id} \rangle \ , 
\qquad S_{\text{pro},j}(\rho) = - \Tr  \tilde \rho_{\text{pro},j} \ln \tilde \rho_{\text{pro},j} . \label{dualproj}
\eeq
Unlike $\rho_{\text{pro},j}$ the new density matrix $\tilde \rho_{\text{pro},j}$ is in general not a faithful representation of the underlying $\rho$ for $j < l$, but the entropy is precisely the same, while its computation involves smaller matrices and is faster. The computational advantage is particularly large for small $j$. 
Expanding the unit operator on $[j]$ in equation (\ref{mixproj}) in terms of basis states, it can be seen that the map $\rho \rightarrow \rho_{\text{pro},j}$ is in fact a trace preserving completely positive map (quantum channel) $[l] \rightarrow [l+j]$ in Kraus form:
\beq
\rho_{\text{pro},j} = \sum_M A_M \rho A_M{}^\dagger \ , \quad \sum A_M{}^\dagger A_M = 1 \ , \quad A_M = \sqrt{\frac{2l+1}{2(l+j)+1}}\Ps_{l+j} |j,M\rangle .
\eeq
There is a similar formula for the transformation of the density matrix in the the dual Gram matrix formulation. In view of the $j\rightarrow \infty$ limit, the lower symbol of a density matrix can also be interpreted as resulting from a completely positive map.

\subsection{$SU(N)$ coherent states and the Lieb-Solovej proof}

Building on the results of the previous section, we will now sketch a few remaining steps in the Lieb-Solovej proof of the entropy conjecture for $SU(N)$,  with the conjecture for $SU(2)$ being a special case. We shall focus on the symmetric representations of $SU(N)$ that act irreducibly on the Hilbert space $\mathcal H_M = \otimes ^M_\text{sym} \mathbb C^N$ of $M$ bosons with $N$ degrees of freedom. The corresponding Young tableaux have a single row. In the spin case  ($N =2$) all irreducible representations are of this form, but for higher $N$ there are other irreps. 
States in the symmetric representations of $SU(N)$ are defined in analogy to the stellar representation~\eqref{stelarrep} 
as the projection $\PS$ onto the fully symmetric part of the tensor product of $M$ unit vectors $\omega_i$ in $\mathbb C^N$
\beq 
|\psi \rangle = c_\psi \PS |\omega_1 \otimes \ldots \otimes \omega_{M}\rangle ,
\eeq
where $c_\psi$ is a normalization constant.\footnote{In this section we use the generic notation $\PS$ for all fully symmetric projectors; their dimensionality follows from context.}
The unit vectors are defined up to a phase, i.e.\  they are really elements of $\mathbb {CP}^{N-1}$, which generalizes the Bloch sphere of the $N=2$ spin case (and is not a sphere for $N>2$). Coherent states in this representation are elements of the $SU(N)$ orbit of a highest weight vector. They are pure condensates of the form $\ket {\Omega_M} = \ket{\Omega \otimes \ldots \otimes \Omega}$ labeled by $\Omega \in \mathbb {CP}^{N-1}$. Everything that we have discussed in the previous sections generalizes to the present case (including the proof for integer R\'enyi entropies). As for spin-coherent states~\eqref{complete}, $\PS$ can be written in terms of a coherent-state POVM. 

Lieb and Solovej prove the following theorem: For all states $\rho$ on $\mathcal H_M$, the ordered eigenvalues of the output of the covariant quantum channel
\beq
\Phi^k(\rho) = \PS (\rho \otimes \mathbf 1_{\otimes^k \mathbb C^N} ) \PS  \label{cloning}
\eeq
are majorized by those of $\Phi^k(|\Omega\rangle\!\langle \Omega|)$, i.e.\ the extremizers are coherent states. This quantum channel is a straightforward generalization of~\eqref{projmatrix}. Interestingly, it has also been introduced as a universal quantum cloning channel~\cite{PhysRevLett.79.2153,PhysRevA.58.1827}: The no-cloning theorem forbids exact copies of a state $\rho$, but one can try to obtain approximate clones with maximum fidelity, meaning that the reduced density matrices of the clones should have maximum overlap with the original state. The quantum channel \eqref{cloning} achieves that in a surprisingly  simple way. The starting point is $\rho \otimes \mathbf 1_{\otimes^k \mathbb C^N}$, which features one perfect copy of $\rho$ and $k$ totally mixed states  (up to normalization), i.e.\ worst possible but universal copies of $\rho$. The expression is then symmetrized with $\PS$ to democratically distribute the original state over all copies. This universal cloning channel has been proven to be optimal in~\cite{PhysRevA.58.1827}.

In the limit $k \rightarrow \infty$, the eigenvalues of \eqref{cloning} approach the values of the lower symbol $\rho(\Omega)$ as we have explained in the previous section. The original entropy conjecture then follows because Shannon entropy is a Schur-concave function of the density $\rho$. Since the proof is obtained using a limit, it does not show uniqueness of the extremizers. 
For the proof, a second quantized formulation in Fock space $\bigotimes_{M=0}^\infty \mathcal H_M$ is convenient: Creation operators $a^*_\omega$ are defined on states $\ket \psi \in \mathcal H_M$ via 
\beq
a^*_\omega \ket\psi = \sqrt{M+1} \PS (\ket \omega \otimes \ket \psi) , 
\eeq
annihilation operators $a_\omega$ are the adjoints of these. For a suitable orthonormal basis $\{\omega_i\}$ we set $a^*_i \equiv a^*_{\omega_i}$ and likewise for $a_i$. These operators satisfy the usual canonical commutation relations $[a_i , a^*_j] = \delta_{ij}$. Lie algebra generators and projection operators find elegant expressions in second quantization via the ``Schwinger trick''. The quantum channel \eqref{original} can be rewritten in second quantized formulation as 
\beq
\Phi^k(\rho) = ... \sum  a^*_{i_1} \cdots a^*_{i_k} \, \rho \, a_{i_k} \cdots a_{i_1} .
\label{original}
\eeq

The starting point of the proof is the $SU(N)$ analog of the covariant quantum channel in the dual Gram picture~\eqref{dualproj}
\beq
\tilde \Phi^k(|\psi\rangle\!\langle \psi|) = \langle \psi \otimes \text{id}_{\otimes^k \mathbb C^N} | \PS | \psi \otimes \text{id}_{\otimes^k \mathbb C^N}\rangle .
\eeq
This quantum channel is known as the universal measure-and-prepare channel in quantum information theory~\cite{10.1007/978-3-642-18073-6_2}. It can be
rewritten in second quantized formulation as
\beq
\tilde \Phi^k(|\psi\rangle\!\langle \psi|) =  \frac 1 {k!} \sum \langle\psi| a_{i_1} \cdots a_{i_k} a^*_{j_k} \cdots a^*_{j_1}|\psi\rangle \, a^*_{i_1} \cdots a^*_{i_k}\, a_{j_k} \cdots a_{j_1} .
\eeq
Now the brilliant idea is to realize that by normal ordering inside the expectation value, this expression can be rewritten in terms of reduced density matrices
\beq
\gamma^l(|\psi\rangle\!\langle \psi|) = \frac 1{l!} \sum  \langle\psi| a^*_{j_l} \cdots a^*_{j_1} a_{i_1} \cdots a_{i_l} |\psi\rangle \, a^*_{i_1} \cdots a^*_{i_l}\, a_{j_l} \cdots a_{j_1} ,
\eeq
and the original quantum channel \eqref{original}, but with lower $k$,
i.e. a proof by induction on $k$ is possible! Indeed
\beq
\tilde \Phi^k(|\psi\rangle\!\langle \psi|) =  \sum_{l=0}^k C_l \,\Phi^l\big(\gamma^{k-l}(|\psi\rangle\!\langle \psi|)\big)
\eeq
with coefficients $C_l$ that are positive and independent of $\psi$, because they simply result from the positive $[a_i, a^*_j]=\delta_{ij}$ commutators. 
 For coherent states the reduced density matrix is again a coherent state (up to normalization) given by
\beq
\gamma^l(|\Omega\rangle\!\langle \Omega|) = \frac{M!}{(M-l)!} |\Omega_{l}\rangle\!\langle \Omega_{l}|
= \frac{M!}{(M-l)! l!} \big(a^*_\Omega\big)^l \big(a_\Omega\big)^l .
\eeq
The majorization theorem follows by induction on $k$. The induction start is $\Phi^0 = \text{id}$. The induction step uses the induction hypothesis, namely
\beq
\Phi^l\big(\gamma^{k-l}(|\Omega\rangle\!\langle \Omega|)\big) = \frac{M!}{(M-k+l)!}  \Phi^l\big(|\Omega_{k-l}\rangle\!\langle \Omega_{k-l}|\big)
\eeq
majorizes $ \Phi^l\big(\gamma^{k-l}(|\psi\rangle\!\langle \psi|)\big)$ for all $k < l$, but the case $k=l$ is obvious, because then $\gamma^{k-l}(|\psi\rangle\!\langle \psi|)=\gamma^0(|\psi\rangle\!\langle \psi|) = 1$ independently of $\psi$.
For further details, we refer to the original paper~\cite{lieb2016proof} and also to~\cite{10.1007/978-3-642-18073-6_2}, where a similar method was used in the context of universal quantum cloning channels.

The investigation of further generalizations of the conjectures and proofs that we have discussed in this article is an active field of research. Quite recently, Lieb and Solovej have made some progress for the interesting case of coherent states of $SU(1,1)$ and its $AX+B$ subgroup, showing in particular that the conjecture holds for integer R\'enyi entropies in the latter case~\cite{lieb2019wehrl}. The whole topic is obviously interesting from a physics point of view (statistical physics, quantum mechanics) as well as from a quantum information perspective. The entropies that we have studied can also be useful tools in statistical data analysis (see e.g.~\cite{PhysRevD.99.103501}).  Furthermore, there seems to be some very interesting mathematics going on that goes beyond what is currently known about convexity in Lie theory~\cite{Neeb2000}.

Let us conclude with a remark on the  angular entropy~\eqref{angentropy}: It was introduced as a toy model for the understanding of the projection and Wehrl entropies~\cite{schupp2008}. For $SU(2)$, its computation involves only $3 \times 3$ matrices and their eigenvalues, but the model easily generalizes to other groups. Theoretical arguments and numerical experiments show that angular entropy shows similar behavior as  Wehrl entropy
and there are also similar conjectures for entropy minimizing states. We shall not resolve this new conjecture here, but suggest it as a nice exercise for the reader.



\begin{ack}
I would like to thank Elliott Lieb for introducing me to this fascinating topic and for many valuable discussions. 
\end{ack}




\bibliographystyle{emss}
\bibliography{Schupp-Wehrl_Entropy}

\begin{thebibliography}{10}
\providecommand{\url}[1]{\texttt{#1}}
\providecommand{\urlprefix}{URL }
\providecommand{\eprint}[2][]{\url{#2}}

\bibitem{Beckner:1975}
W.~Beckner, Inequalities in {Fourier} analysis. \emph{Ann. Math.} \textbf{102}
  (1940), 264–314

\bibitem{bengtsson_zyczkowski_2017}
I.~Bengtsson and K.~Życzkowski, \emph{Geometry of quantum states: An
  introduction to quantum entanglement}. 2nd edn., Cambridge University Press,
  2017

\bibitem{Bodman2004}
B.~G. Bodmann, A lower bound for the {Wehrl} entropy of quantum spin with sharp
  high-spin asymptotics. 2004

\bibitem{BRASCAMP1976151}
H.~J. Brascamp and E.~H. Lieb, Best constants in {Young's} inequality, its
  converse, and its generalization to more than three functions. \emph{Advances
  in Mathematics} \textbf{20} (1976), 151--173

\bibitem{Carlen1991SomeII}
E.~A. Carlen, Some integral identities and inequalities for entire functions
  and their application to the coherent state transform. \emph{J. Funct.
  Analysis} \textbf{97} (1991), 231--249

\bibitem{10.1007/978-3-642-18073-6_2}
G.~Chiribella, On quantum estimation, quantum cloning and finite quantum de
  finetti theorems. In \emph{Theory of quantum computation, communication, and
  cryptography}, edited by W.~van Dam, V.~M. Kendon, and S.~Severini, pp.
  9--25, Springer Berlin Heidelberg, Berlin, Heidelberg, 2011

\bibitem{DePalma2018}
G.~De~Palma, The {Wehrl} entropy has {Gaussian} optimizers. \emph{Lett. Math.
  Phys.} \textbf{108} (2018), 97–116

\bibitem{PhysRevLett.79.2153}
N.~Gisin and S.~Massar, Optimal quantum cloning machines. \emph{Phys. Rev.
  Lett.} \textbf{79} (1997), 2153--2156

\bibitem{Helling:2006xh}
R.~C. Helling, P.~Schupp, and T.~Tesileanu, {{CMB} statistical anisotropy,
  multipole vectors and the influence of the dipole}. \emph{Phys. Rev. D}
  \textbf{74} (2006), 063004

\bibitem{Husimi:1940}
K.~Husimi, Some formal properties of the density matrix. \emph{Proc. Phys.
  Math. Soc. Jpn.} \textbf{22} (1975), 159--182

\bibitem{Lee_1988}
C.~T. Lee, Wehrl's entropy of spin states and {Lieb's} conjecture. \emph{J.
  Phys. A} \textbf{21} (1988), 3749--3761

\bibitem{Liebconvexity:1975}
E.~Lieb, Some convexity and subadditivity properties of entropy. \emph{Bull.
  Am. Math. Soc.} \textbf{81} (1975), 1--13

\bibitem{lieb2019wehrl}
E.~Lieb and J.~Solovej, Wehrl-type coherent state entropy inequalities for
  {SU(1,1)} and its {AX+B} subgroup. In \emph{Partial differential equations,
  spectral theory, and mathematical physics}, pp. 301--314, EMS, 2021

\bibitem{Lieb:1978ys}
E.~H. Lieb, {Proof of an Entropy Conjecture of Wehrl}. \emph{Commun. Math.
  Phys.} \textbf{62} (1978), 35

\bibitem{Lieb:1991np}
E.~H. Lieb and J.~P. Solovej, {Quantum coherent operators: A Generalization of
  coherent states}. \emph{Lett. Math. Phys.} \textbf{22} (1991), 145--154

\bibitem{lieb2014proof}
E.~H. Lieb and J.~P. Solovej, Proof of an entropy conjecture for {Bloch}
  coherent spin states and its generalizations. \emph{Acta Mathematica}
  \textbf{212} (2014), no.~2, 379--398

\bibitem{lieb2016proof}
E.~H. Lieb and J.~P. Solovej, Proof of the {Wehrl}-type entropy conjecture for
  symmetric {SU(N)} coherent states. \emph{Commun. Math. Phys.} \textbf{348}
  (2016), 567--578

\bibitem{Luo_2000}
S.~Luo, A simple proof of {Wehrl's} conjecture on entropy. \emph{J. Phys. A}
  \textbf{33} (2000), 3093--3096

\bibitem{PhysRevD.99.103501}
M.~Minkov, M.~Pinkwart, and P.~Schupp, Entropy methods for {CMB} analysis of
  anisotropy and non-{Gaussianity}. \emph{Phys. Rev. D} \textbf{99} (2019),
  103501

\bibitem{Neeb2000}
K.-H. Neeb, \emph{{Holomorphy} and convexity in {Lie} theory}. Expositions in
  Mathematics 28, Walter de Gruyter \& Co., Berlin, 2000

\bibitem{Perelomov:book}
A.~Perelomov, \emph{Generalized coherent states and their applications}.
  Springer-Verlag, Berlin, 1986

\bibitem{1926Schroedinger}
E.~Schrödinger, {Der stetige Übergang von der Mikro- zur Makromechanik}.
  \emph{Naturwissenschaften} \textbf{14} (1926), no.~28, 664--666

\bibitem{Schupp1999OnLC}
P.~Schupp, On {Lieb's} conjecture for the {Wehrl} entropy of {Bloch} coherent
  states. \emph{Commun. Math. Phys.} \textbf{207} (1999), 481--493

\bibitem{SchuppKolloquium2004}
P.~Schupp, {Zu Liebs Vermutung über die Wehrl Entropie von Quantenspins}.
  {Kolloquium über Reine Mathematik, Universität Hamburg}, 2004

\bibitem{schupp2008}
P.~Schupp. unpublished, 2008

\bibitem{Horia2002}
H.~Scutaru, On {Lieb's} conjecture. \emph{Romanian J. of Phys.} \textbf{47}
  (2002), 189--198, preprint: {FT--180--1979}

\bibitem{Wehrl:1978zz}
A.~Wehrl, {General properties of entropy}. \emph{Rev. Mod. Phys.} \textbf{50}
  (1978), 221--260

\bibitem{Wehrl:1979}
A.~Wehrl, On the relation between classical and quantum-mechanical entropy.
  \emph{Rept. Math. Phys.} \textbf{16} (1979), 353--358

\bibitem{Weingart}
G.~Weingart, {Explicit computation of symmetric polynominals in the expression
  for the Wehrl entropy at higher spin}. personal communication, unpublished,
  2004

\bibitem{PhysRevA.58.1827}
R.~F. Werner, Optimal cloning of pure states. \emph{Phys. Rev. A} \textbf{58}
  (1998), 1827--1832

\end{thebibliography}

\end{document}